\documentclass[letter]{jpsj3} 
\title{Weak superconducting fluctuations and small anisotropy of the upper critical fields in an Fe$_{1.05}$Te$_{0.85}$Se$_{0.15}$ single crystal}

\author{
Takanori \textsc{Kida}$^{1,4}$\thanks{E-mail address: kida@mag.cqst.osaka-u.ac.jp}, 
Masahiro \textsc{Kotani}$^{1}$, 
Yoshikazu \textsc{Mizuguchi}$^{2,3,4}$, \\
Yoshihiko \textsc{Takano}$^{2,3,4}$, 
and Masayuki \textsc{Hagiwara}$^{1,4}$
}

\inst{$^{1}$KYOKUGEN, Osaka University, 1-3 Machikaneyama, Toyonaka, Osaka 560-8531, Japan\\
$^{2}$National Institute for Materials Science, 1-2-1 Sengen, Tsukuba 305-0047, Japan\\
$^{3}$University of Tsukuba, 1-1-1 Tennodai, Tsukuba 305-8577, Japan\\
$^{4}$JST, Transformative Research-Project on Iron Pnictides (TRIP), Chiyoda-ku, Tokyo 102-0075, Japan
}

\abst{
We have investigated the temperature dependence of the resistive upper critical fields ($\mu_{0}H_{\rm c2}(T)$) for an Fe$_{1.05}$Te$_{0.85}$Se$_{0.15}$ single crystal, which exhibits superconductivity at $T_{\rm c} \sim 14$~K, in magnetic fields of up to 55~T. 
Two-dimensional feature and superconducting fluctuations of the samples are found to be weak, because the resistive broadening effect on applied magnetic fields of up to 14~T is small. 
The Pauli paramagnetic effect is obviously evidenced by the strong suppression of the $\mu_{0}H_{\rm c2}^{ab}(T)$ ($H \parallel ab$) curve and nearly isotropic $\mu_{0}H_{\rm c2}(0) \approx 47$~T is seen for both $H \parallel ab$ and $H \parallel c$. 
This fact is almost identical to the results of Fe$_{1+y}$Te$_{0.6}$Se$_{0.4}$ single crystals reported previously. 
We have discussed that the small anisotropy of the upper critical field at low temperatures in Fe$_{1+y}$(Te,Se) systems against the variation of the Te/Se ratio.
} 

\kword{iron-based superconductor, Fe$_{1+y}$(Te,Se), upper critical field, anisotropy}

\begin{document}
\maketitle

The discovery of high-temperature superconductivity in the iron-pnictide LaFeAsO$_{1-x}$F$_{x}$~\cite{Kamihara} 
(abbreviated as the 1111-system) and related compounds has generated great interest in understanding 
the interplay of magnetism, superconductivity, and electrical structure. 
So far, several other groups of iron-based superconductors have been discovered, 
such as $Ae$Fe$_{2}$As$_{2}$ (abbreviated as the 122-system, $Ae$~=~alkali earth metals)~\cite{Rotter}, 
LiFeAs (abbreviated as the 111-system)~\cite{Pitcher}, 
and tetragonal Fe$Ch$ (abbreviated as the 11-system, $Ch$~=~chalcogen).~\cite{Hsu} 
These iron-based compounds display a phase transition from antiferromagnetic to superconducting ground states 
tuned by chemical doping~\cite{Kamihara} or external pressure,~\cite{Takahashi} 
suggesting that the antiferromagnetic spin fluctuations of Fe play an important role in developing 
the superconducting ground states~\cite{Rotter}. 
The 11-system superconductors, such as Fe$_{1+\delta}$Se~\cite{Hsu,Mizuguchi1}, FeTe$_{1-x}$Se$_{x}$~\cite{Yeh}, and FeTe$_{1-x}$S$_{x}$~\cite{Mizuguchi2}, are of great importance 
in understanding the mechanism of superconductivity in iron-based superconductors owing to their simple structures. 
Fe$_{1+\delta}$Se, which exhibits superconductivity at $T_{\rm c}=8$~K, 
has a tetragonal PbO-type structure ($P_{4}/nmm$) composed of stacked FeSe layers along 
the $c$-axis~\cite{Hsu}. 
The superconductivity in Fe$_{1+\delta}$Se is significantly affected by external 
pressure~\cite{Mizuguchi1,Masaki,Margadonna,Medvedev} and chalcogenide substitutions~\cite{Yeh,Mizuguchi2}. 
In particular, an applied pressure of only up to 4.15~GPa drastically enhances its $T_{\rm c}$ to $\sim 37$~K (d~ln$T_{\rm c}$/d$P \sim 0.91$)~\cite{Masaki}. 
The pressure effect of superconductivity in Fe$_{1+\delta}$Se is larger than that in other iron-based 
superconductors, $e.g.$, 
the $T_{\rm c}$ of LaFeAsO$_{1-x}$F$_{x}$ increases from 26~K at ambient pressure to 43~K at 4~GP 
(d~ln$T_{\rm c}$/d$P \sim 0.16$)~\cite{Takahashi}. 

For a thorough understanding of the mechanism of superconductivity in iron-based superconductors, 
it is important to study the upper critical field ($\mu_{0}H_{\rm c2}$) because the $\mu_{0}H_{\rm c2}$ provides information on anisotropy, coherent length, effective electron mass, and the pair-breaking mechanism. 
In general, high $T_{\rm c}$ superconductors show extremely high $\mu_{0}H_{\rm c2}$ values. 
The temperature dependence of $\mu_{0}H_{\rm c2}(T)$ in low magnetic fields is often a very poor guide to 
its intrinsic features at low temperatures. 
Therefore, transport measurements in very high magnetic fields and at low temperatures close to $T=0$~K 
have provided useful information not only on $\mu_{0}H_{\rm c2}(0)$ but also on the nature of 
the phase in the vicinity of a quantum phase transition point. 
Previously, we reported the temperature dependence of $\mu_{0}H_{\rm c2}(T)$ on FeTe$_{0.75}$Se$_{0.25}$ polycrystals~\cite{Kida}, which showed a strong suppression effect at low temperatures 
due to the Pauli paramagnetic effect. 
In the present study, we performed electrical resistivity measurements of a single crystal of Fe$_{1.05}$Te$_{0.85}$Se$_{0.15}$ in magnetic fields of up to 55~T 
to discuss the anisotropy of $\mu_{0}H_{\rm c2}(T)$. 
Recently, Fang $et~al.$~\cite{Fang} and Khim $et~al.$~\cite{Khim} reported the temperature dependence of 
$\mu_{0}H_{\rm c2}(T)$ of Fe$_{1+y}$Te$_{0.6}$Se$_{0.4}$ single crystals. 
Therefore, we compare and discuss the $\mu_{0}H_{\rm c2}(T)$s on these Fe$_{1+y}$(Te,Se) systems.


Single crystals of Fe$_{1+y}$Te$_{1-x}$Se$_{x}$ were grown with a self-flux method as described in ref.~\ref{sample}. 
After annealing, the present single crystal of Fe$_{1.05}$Te$_{0.85}$Se$_{0.15}$, whose composition was determined by 
the energy dispersive x-ray spectroscopy analysis, was obtained from a precursor material with nominal 
composition FeTe$_{0.75}$Se$_{0.25}$. 
We prepared a sample with typical dimensions of 
800$\times$200$\times$15~$\mu$m$^{3}$ for electrical resistivity ($\rho$) measurements. 
The sample is mounted on a sapphire substrate (4$\times$0.5$\times$20~mm$^{3}$), which is a good 
thermal anchor. 
The temperature dependence of $\rho(H)$ 
was measured by a dc four-probe technique in static magnetic fields of up to 14~T with a commercial 
superconducting magnet system (Oxford Instruments Ltd.). 
The electrical current direction was parallel to the $ab$-plane of the sample, and 
the magnetic field was applied along the $ab$-plane or the $c$-axis. 
The $\rho(H)$ in pulsed magnetic fields of up to 55~T was measured by utilizing a non-destructive pulsed 
magnet. 
The duration of the pulsed magnetic field was about 40~msec.

\begin{figure}[tp]
\begin{center}
\includegraphics[width=0.6\textwidth,keepaspectratio=true]{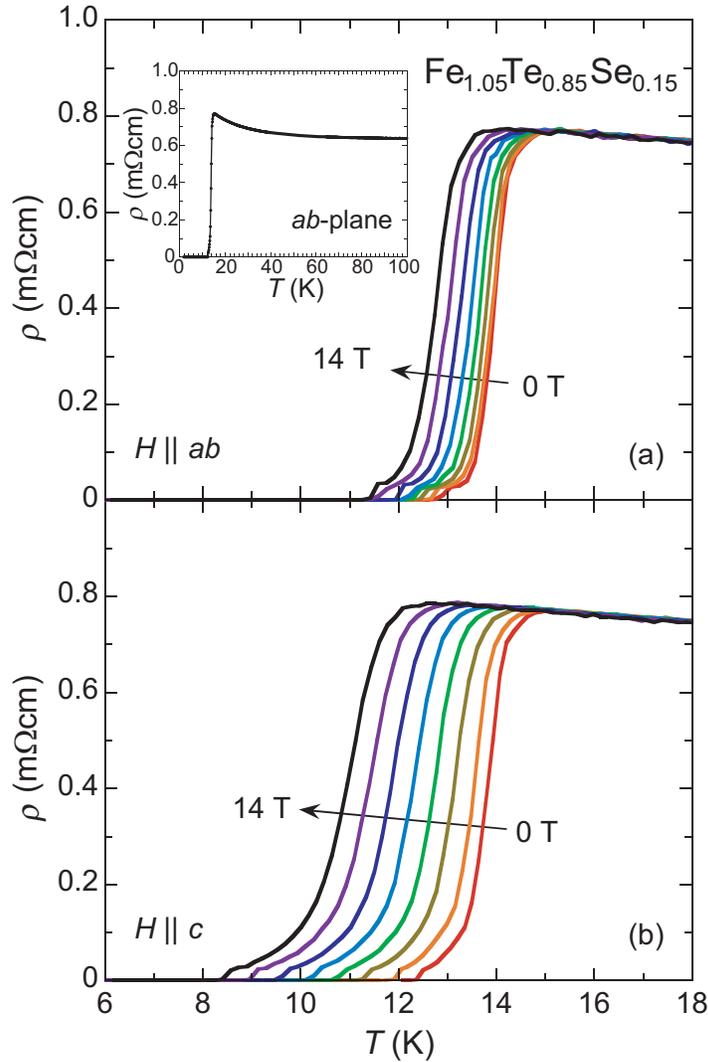}
\end{center}
\caption{(Color online)~Temperature dependence of the electrical resistivity of a single crystal of Fe$_{1.05}$Te$_{0.85}$Se$_{0.15}$ 
in static magnetic fields of up to 14~T with increments of 2~T for (a)~$H \parallel ab$ and (b)~$H \parallel c$. 
The inset displays the temperature dependences of the zero-field resistivity in the $ab$-plane below 100~K.
}
\label{fig01}
\end{figure}
Figures~\ref{fig01}(a) and \ref{fig01}(b) show the temperature dependence of the $\rho(H)$ 
in static magnetic fields of up to 14~T with increments of 2~T for $H \parallel ab$ and $H \parallel c$, respectively. 
The zero-field resistivity exhibits a superconducting transition 
at $T_{\rm c}^{\rm onset} = 14.1$~K as shown in the inset of Fig.~\ref{fig01}(a). 
Above $T_{\rm c}$, the zero-field resistivity of the present sample increases with decreasing temperature. 
Similar behavior was reported in the literature~\cite{Liu,Fang} on the transport properties of Fe$_{1.11}$Te$_{0.6}$Se$_{0.4}$ single crystals. 
This ``semiconducting" behavior has been attributed to a weak charge-carrier localization due to 
a large amount of excess Fe in Fe$_{1+y}$Te$_{1-x}$Se$_{x}$ systems~\cite{Liu}. 

For the present sample, $T_{\rm c}$ decreases with increasing magnetic fields for both $H \parallel ab$ and $H \parallel c$. 
The shifts of $T_{\rm c}^{\rm onset}$ for $H \parallel ab$ and $H \parallel c$ at 14~T from the zero field value are 
$\Delta T_{\rm c}^{ab}=-1.5$~K and $\Delta T_{\rm c}^{c}=-3.2$~K, respectively. 
The $\rho(H)-T$ curves for both $H \parallel ab$ and $H \parallel c$ shift parallel to the low temperature side with 
increasing magnetic fields without broadening, 
suggesting that the two-dimensional feature of the 
present sample is smaller than that of high-$T_{\rm c}$ cuprates~\cite{Ikeda}. 
In general, the resistive-broadening effect of high-$T_{\rm c}$ cuprates is attributed to strong superconducting 
fluctuations due to the high two-dimensionality of the CuO$_{2}$-plane~\cite{Ikeda}. 
In our results, however, a weak resistive-broadening is observed at higher magnetic fields for $H \parallel c$, 
which probably shows an effect from the vortex motion~\cite{Tinkham} as well as reported results for iron-chalcogenide 
superconductors~\cite{Liu,Fang} and other iron-pnictide ones.~\cite{Jia}

\begin{figure*}[bp]
\begin{center}
\includegraphics[width=1\textwidth,keepaspectratio=true]{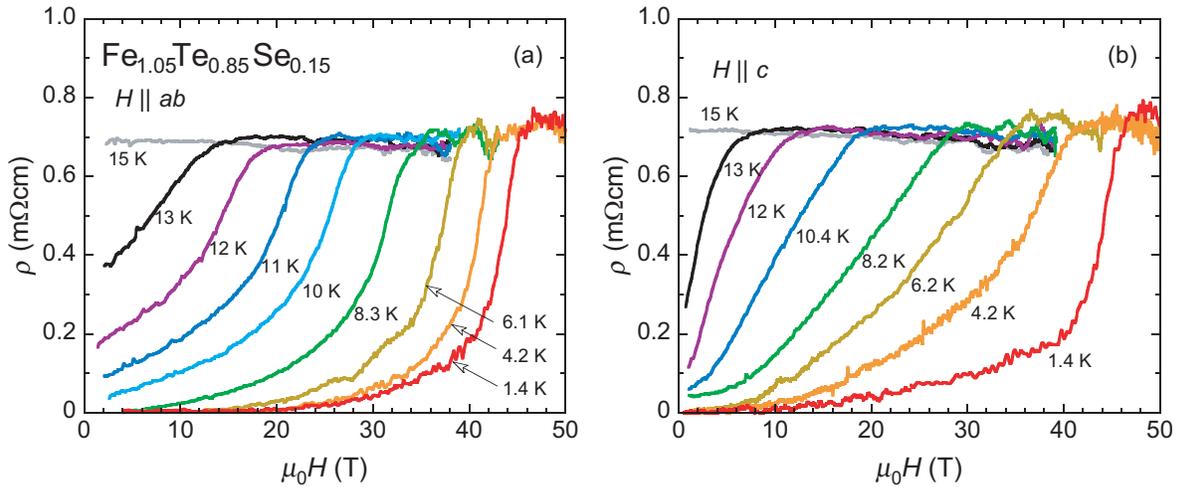}
\end{center}
\caption{(Color online)~Electrical resistivity $\rho(H)$ as a function of magnetic field of up to 50~T at designated temperatures for (a)~$H \parallel ab$ and (b)~$H \parallel c$.}
\label{fig02}
\end{figure*}
Figures~\ref{fig02}(a) and \ref{fig02}(b) depict 
the magnetic field dependence of the resistivities in the $ab$-plane at designated temperatures for 
$H \parallel ab$ and $H \parallel c$, respectively. 
The resistivities show no hysteresis between the field ascending and descending processes, 
indicating no measurable heating due to eddy currents induced 
in the sample caused by the pulsed magnetic field. 
Upon heating from the lowest temperature, 
the superconducting to normal state transitions shift to lower magnetic fields. 
The transition induced by applied magnetic fields is considerably broad. 
At $T=1.4$~K, 
the values of the zero-resistivity field $\mu_{0}H_{\rm c2}^{\rm zero}$ (10~\% of the normal state resistivity) for 
$H \parallel ab$ and $H \parallel c$ are $\sim 34$~T and $\sim 26$~T, respectively. 
It is expected that the appearance of finite resistivity with lowering magnetic field for $H \parallel c$ 
arises from dissipation associated with thermally activated vortex motion.~\cite{Tinkham} 
The onset upper critical fields $\mu_{0}H_{\rm c2}^{\rm onset}$ 
(90~\% of the normal state resistivity) at 1.4~K for $H \parallel ab$ and $H \parallel c$ are almost equal; 
$\mu_{0}H_{\rm c2}^{ab}=45$~T and $\mu_{0}H_{\rm c2}^{c}=46$~T. 
Similar results were reported by Fang $et~al.$~\cite{Fang} and Khim $et~al.$~\cite{Khim} 
for Fe$_{1+y}$Te$_{0.6}$Se$_{0.4}$ single crystals.

\begin{figure}[tp]
\begin{center}
\includegraphics[width=0.6\textwidth,keepaspectratio=true]{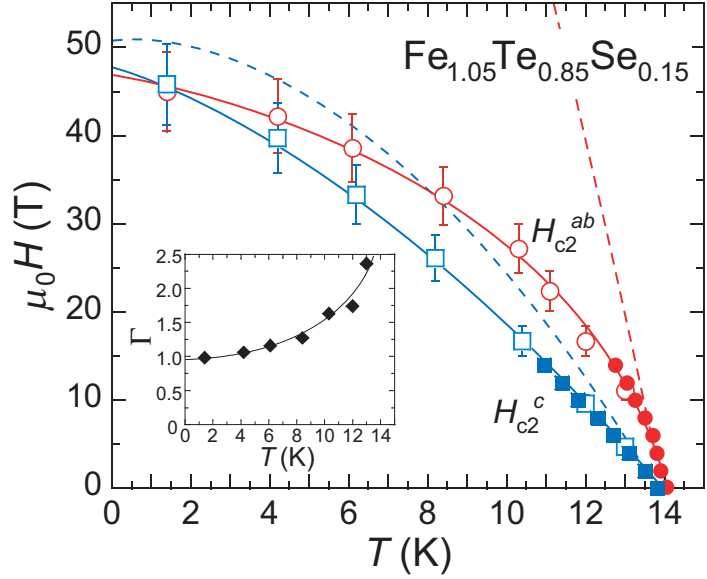}
\end{center}
\caption{(Color online)~Field-temperature ($H$-$T$) phase diagram for a single crystal of Fe$_{1.05}$Te$_{0.85}$Se$_{0.15}$ for $H \parallel ab$ ($H_{\rm c2}^{ab}$) and $H \parallel c$ ($H_{\rm c2}^{c}$). The solid and open symbols correspond to the data in the static and pulsed magnetic fields, respectively. The broken lines represent the WHH prediction with only the orbital pair-breaking effect. The solid lines are added for eye-guides.}
\label{fig03}
\end{figure}
Using the magnetic field and temperature dependences of the resistivity in static and pulsed magnetic fields, 
we show the field-temperature ($H$-$T$) phase diagram of the present sample in Fig.~\ref{fig03}. 
The onset values of $\mu_{0}H_{\rm c2}(T)$ and $T_{\rm c}(H)$ in the superconducting transitions were plotted. 
The solid and open symbols in Fig.~\ref{fig03} correspond to the data in the static and the pulsed magnetic fields, 
respectively. 
The slopes (${\rm d}\mu_{0}H_{\rm c2}/{\rm d}T$) for $H \parallel ab$ and $H \parallel c$ 
at $T_{\rm c}^{\rm onset}(0)$ were $-17.2$ and $-5.2$~T/K, respectively, which were estimated by fitting 
the $T_{\rm c}^{\rm onset}(H)$ data in the static magnetic fields up to 8~T to a linear function. 
The isotropically averaged value of ${\rm d}\mu_{0}H_{\rm c2}/{\rm d}T~(= -13.2~$T/K) for the single crystal is close to 
the FeTe$_{0.75}$Se$_{0.25}$ polycrystal's value ($-13.7$~T/K), which was reported by us previously.~\cite{Kida} 
On the basis of the conventional one-band Werthamer-Helfand-Hohenberg (WHH) theory~\cite{WHH}, 
$i.e.$, the dirty-limit Bardeen-Cooper-Schrieffer (BCS) relation of 
$\mu_{0}H_{\rm c2}(0)=-0.69T_{\rm c}({\rm d}\mu_{0}H_{\rm c2}/{\rm d}T)|_{T_{\rm c}}$, 
we can obtain the orbital pair-breaking fields ($\mu_{0}H_{\rm c2}^{\rm orb}(0)$), 
$167$~T for $H \parallel ab$ and $50.3$~T for $H \parallel c$. 
The temperature dependence of the upper critical field empirically obeys the relation 
$\mu_{0}H_{\rm c2}(T)/ \mu_{0}H_{\rm c2}(0) = 1-(T/T_{\rm c})^{2}$. 
The broken lines in Fig.~\ref{fig03} represent the WHH prediction with only the orbital pair-breaking effect. 
Experimental curves of $\mu_{0}H_{\rm c2}^{ab}(T)$ and $\mu_{0}H_{\rm c2}^{c}(T)$ are suppressed at low temperatures 
compared to the WHH lines. 
Especially, the $\mu_{0}H_{\rm c2}^{ab}(T)$ is considerably small at low temperatures, suggesting that 
the Pauli paramagnetic effect (called also the spin Zeeman effect) is obviously dominant for the pair-breaking 
as well as the result of the FeTe$_{0.75}$Se$_{0.25}$ polycrystal sample.~\cite{Kida} 
At $T \rightarrow 0$, both $\mu_{0}H_{\rm c2}^{ab}(T)$ and $\mu_{0}H_{\rm c2}^{c}(T)$ are extrapolated to 
$\sim 47$~T, which is consistent with the value determined for the FeTe$_{0.75}$Se$_{0.25}$ 
polycrystal~\cite{Kida}. 
Near $T_{\rm c}$, the ${\rm d}\mu_{0}H_{\rm c2}^{c}/{\rm d}T$ is almost linear, but 
the ${\rm d}\mu_{0}H_{\rm c2}^{ab}/{\rm d}T$ seems to be nonlinear. 
This indicates that the suppression of the upper critical field for $H \parallel ab$ occurs not only at low temperatures but also near $T_{\rm c}$. 
Similar results have been observed in a spin-triplet superconductor Sr$_{2}$RuO$_{4}$,~\cite{Kittaka} in which 
the remarkable suppression of the upper critical field appears for $H \parallel [100]$, and the origin of the 
suppression remains to be clarified. 
The temperature dependence of the anisotropy coefficient $\Gamma(T)$ defined by 
$H_{\rm c2}^{ab}(T)/H_{\rm c2}^{c}(T)$ is shown in the inset of Fig.~\ref{fig03}. 
The $\Gamma(T)$ decreases from $\sim 2.4$ near $T_{\rm c}$ to $\sim 1$ at $T=0$, monotonically. 
The $\mu_{0}H_{\rm c2}^{ab}(T)$, $\mu_{0}H_{\rm c2}^{c}(T)$, and $\Gamma(T)$ of 
Fe$_{1.05}$Te$_{0.85}$Se$_{0.15}$ are almost equal to those of Fe$_{1+y}$Te$_{0.6}$Se$_{0.4}$,~\cite{Fang,Khim} 
indicating that the small anisotropy of the upper critical field at low temperatures is robust against 
the variation of the Te/Se ratio, at least in these samples. 
Similar isotropic behavior of the upper critical field at low temperatures has also been observed in the 122-system 
of iron-based superconductors.~\cite{Yuan,Yamamoto}
These results indicate that the small anisotropy of the upper critical field at low temperatures may be 
a general feature on the 11-system and the 122-system of iron-based superconductors, 
being consistent with band calculations~\cite{band1,band2} and angle resolved photoemission spectroscopy (ARPES) results~\cite{ARPES1}. 

From the Ginzburg-Landau (G-L) theory~\cite{Tinkham}, 
the coherent length tensor $\xi(0)$ is calculated from the extrapolated $\mu_{0}H_{\rm c2}^{ab}(0)$ and 
$\mu_{0}H_{\rm c2}^{c}(0)$ data using the relations given by 
$\mu_{0}H_{\rm c2}^{ab}(0)=\Phi_{0}/2\pi \xi_{ab}(0)\xi_{c}(0)$, 
$\mu_{0}H_{\rm c2}^{c}(0)=\Phi_{0}/2\pi \xi_{ab}^{2}(0)$, 
where $\Phi_{0}=2\pi \hbar/2e \simeq 2.0678$~Tm$^{2}$ is the flux quantum. 
The obtained values are $\xi_{ab}(0) \approx \xi_{c}(0) \approx 26$~\AA. 
Near $T_{\rm c}$ ($T<T_{\rm c}$), the $\mu_{0}H_{\rm c2}(T)$ and the $\xi(T)$ are phenomenologically proportional to 
$1-(T/T_{\rm c})^{2}$ and $(1-T/T_{\rm c})^{-1/2}$, respectively. 
Considering the temperature dependence of $\Gamma(T)$ and the strong suppression of $\mu_{0}H_{\rm c2}^{ab}(T)$, 
it is expected that $\xi_{c}(T)$ is much smaller than $\xi_{ab}(T)$ near $T_{\rm c}$. 
The small $\xi_{c}(T)$ seems to correlate with the fact that $\Delta T_{\rm c}^{c}$ is larger than $\Delta T_{\rm c}^{ab}$.

\begin{figure}[bp]
\begin{center}
\includegraphics[width=0.6\textwidth,keepaspectratio=true]{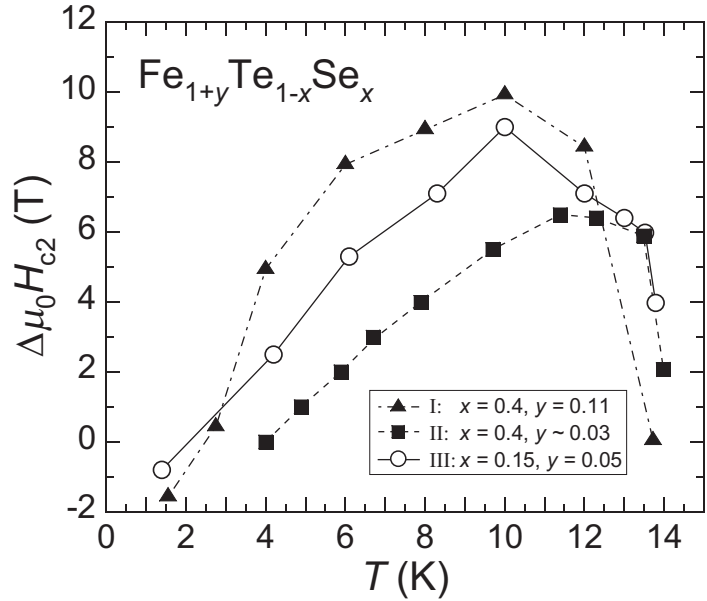}
\end{center}
\caption{Temperature dependence of the $\Delta \mu_{0}H_{\rm c2} (= \mu_{0}[H_{\rm c2}^{ab}-H_{\rm c2}^{c}])$ for $x=0.4,~y=0.11$ in ref.~\ref{Fang}, $x=0.4,~y \sim 0.03$ in ref.~\ref{Khim}, and $x=0.15,~y=0.05$ in the present study.}
\label{fig04}
\end{figure}
Finaly, 
we discuss the composition dependence of the upper critical fields on the Fe$_{1+y}$(Te,Se) systems. 
Figure~\ref{fig04} shows the temperature dependences of 
$\Delta \mu_{0}H_{\rm c2} (= \mu_{0}[H_{\rm c2}^{ab}-H_{\rm c2}^{c}])$ for $x=0.4,~y=0.11$ in ref.~\ref{Fang} (sample~I), $x=0.4,~y \sim 0.03$ in ref.~\ref{Khim} (sample~II), and $x=0.15,~y=0.05$ in the present study (sample III). 
Khim $et~al.$~\cite{Khim} predicts that the sample~II is close to the stoichiometric Fe(Te$_{0.6}$Se$_{0.4}$) with 
a minimal amount of excess and/or interstitial Fe by considering its metallic resistivity above $T_{\rm c}$. 
In comparison with the resistivity results by Liu $et~al.$~\cite{Liu}, 
it is expected that the value of $y$ for sample~II is close to 0.03. 
The $\Delta \mu_{0}H_{\rm c2}$s for all samples first greatly increase with decreasing temperature near $T_{\rm c}$, reach maximum values of $6\sim 10$~T, and then decrease with decreasing temperature. 
The $\Delta \mu_{0}H_{\rm c2}$ of sample~I is larger than that of sample~II in the mid-temperature region below $T_{\rm c}$, indicating that the variation of excess Fe yields a difference in the temperature dependence of 
$\Delta \mu_{0}H_{\rm c2}$. 
The maximum position of $\Delta \mu_{0}H_{\rm c2}$ shifts to the low temperature side with increasing $y$. 
The Te/Se ratio of sample~III is larger than that of sample~II, although the values of $y$ for both samples 
are almost equal. 
The temperature dependence of $\Delta \mu_{0}H_{\rm c2}$ for sample~III is similar to that of sample~I, 
suggesting that the variation of the Te/Se ratio causes no change on the temperature dependence of 
$\Delta \mu_{0}H_{\rm c2}$. 
Both samples~I and III show ``semiconducting" behavior above $T_{\rm c}$. 
It is expected that this behavior is attributable to non-stoichiometry, such as excess Fe and/or defects of chalcogen ions, in Fe$_{1+y}$(Te,Se) systems. 
Using the G-L coherent lengths, the temperature dependence of $\Delta \mu_{0}H_{\rm c2}$ is expressed as follows, 
$\Delta \mu_{0}H_{\rm c2} = \Phi_{0}(\xi_{ab}-\xi_{c})/2\pi \xi_{ab}^{2}\xi_{c}$. 
It may be possible that the $\Delta \mu_{0}H_{\rm c2}$ increases due to the decrease of $\xi_{c}$. 
However, it should be pointed out that those estimations are based on the one-band G-L theory 
which may not be valid for the multi-band compound, and much more information will be needed for further theoretical 
and/or experimental investigations and discussion on superconductivity in this system. 
Nevertheless, these samples show the common feature of 
$\mu_{0}H_{\rm c2}^{ab}(0) \approx \mu_{0}H_{\rm c2}^{c}(0)$ at low temperatures. 
More detailed investigation for the composition dependence of $\mu_{0}H_{\rm c2}(T)$ in Fe$_{1+y}$(Te,Se) systems is necessary to understand this fact.

In conclusion, 
we have performed electrical resistivity measurements on a single crystal of Fe$_{1.05}$Te$_{0.85}$Se$_{0.15}$, which exhibits superconductivity at $T_{\rm c}^{\rm onset}=14.1$~K, in magnetic fields of up to 55~T. 
The $\rho(H)-T$ curves for both $H \parallel ab$ and $H \parallel c$ shift parallel to the low temperature 
side in magnetic fields of up to 14~T, suggesting that the two-dimensional feature and the superconducting fluctuations of the present sample are small. 
The slopes (${\rm d}\mu_{0}H_{\rm c2}/{\rm d}T$) for $H \parallel ab$ and $H \parallel c$ at $T_{\rm c}^{\rm onset}(0)$ are largely different, 
but both $\mu_{0}H_{\rm c2}^{ab}(T)$ and $\mu_{0}H_{\rm c2}^{c}(T)$ are extrapolated to $\sim 47$~T at $T \rightarrow 0$, 
which is consistent with the reported results for Fe$_{1+y}$Te$_{0.6}$Se$_{0.4}$ single crystals.~\cite{Fang,Khim} 
The anisotropy coefficient $\Gamma(T)$ decreases from $\sim 2.4$ near $T_{\rm c}$ to $\sim 1$ at low temperatures. 
The nearly isotropic upper critical field at low temperatures is robust against not only the variation of excess Fe but also that of the Te/Se ratio. 
The observation of the strong suppression in $\mu_{0}H_{\rm c2}^{ab}(T)$ supports the presence of the Pauli paramagnetic effect.

\section*{Acknowledgements} 
This work was partly supported by 
a Grant-in-Aid for ``Transformative Research-project on Iron Pnictides (TRIP)" 
from the Japan Science and Technology Agency (JST), 
Grants-in-Aid for Scientific Research on priority Areas ``High Field Spin Science in 100T" (No.451) 
and ``New Materials Science Using Regulated Nano Spaces" (No.19051010) 
from the Ministry of Education, Culture, Sports, Science and Technology (MEXT), Japan, 
and the Global COE Program 
(Core Research and Engineering of Advanced Materials-Interdisciplinary Education Center for 
Materials Science) (No. G10) from the MEXT, Japan.

\end{document}